\documentclass[twocolumn,prx,aps,amsmath,amssymb,longbibliography]{revtex4-1}

\usepackage{graphicx}
\usepackage{dcolumn}
\usepackage{bm}

\begin{document}

\title{Fluctuation response of a minimal Kitaev chain in nonequilibrium states}

\author{Sergey Smirnov}
\affiliation{P. N. Lebedev Physical Institute of the Russian Academy of
  Sciences, 119991 Moscow, Russia}

\email{1) sergej.physik@gmail.com\\2)
  sergey.smirnov@physik.uni-regensburg.de\\3) ssmirnov@sci.lebedev.ru}

\date{\today}

\begin{abstract}
Minimal Kitaev chains provide a unique platform to engineer Majorana states in
quantum dots interacting via normal tunneling and crossed Andreev reflection
specified by their amplitudes $|\eta_{n,a}|$. Here we analyze fluctuations of
electric currents in a double quantum dot Kitaev chain using the differential
effective charge $q$, that is the ratio of the differential shot noise and
conductance. At low bias voltages $V$ we find that $q=e/2$ in a very narrow
vicinity of the point $|\eta_n|=|\eta_a|$ whereas $q=3e/2$ almost in the whole
sweet spot region and marks the range where the poor man's Majorana states
largely govern the fluctuations. At high $V$ we show that the sweet spot
region is still characterized by $q=3e/2$ uniquely identifying the poor man's
Majorana states using the high voltage tails. For $|\eta_n|=0$ or $|\eta_a|=0$
we obtain $q=e$ at any $V$. Remarkably, before the asymptotic value $q=e$ is
reached for very high $V$, the maximal value $q=2e$ is formed at
$|eV|=2\sqrt{|\eta_n|^2+|\eta_a|^2}$. The unique nature and potentially rich
fluctuation behavior revealed in this work provide a stimulating ground for
the next generation experiments on nonequilibrium shot noise in minimal Kitaev
chains.
\end{abstract}

\maketitle

\section{Introduction}\label{intro}
A complex technological design and comprehensive optimization of various
nanoscale architectures aimed to generate and make non-Abelian Majorana bound
states (MBSs) a practical tool in fault-tolerant quantum computations
\cite{Kitaev_2003} greatly advance our understanding of low-energy behavior
emerging in the corresponding composite superconducting nanostructures. This
is exemplified by extensive research on physical properties of topological
superconductors in diverse hybrid semiconductor(topological
insulator)-superconductor heterostructures \cite{Alicea_2012,Flensberg_2012,
Sato_2016,Aguado_2017,Lutchyn_2018,Marra_2022,Muralidharan_2023,Tanaka_2024}
which represent highly optimized descendants of the paradigmatic physical
realizations \cite{Fu_2008,Fu_2009,Lutchyn_2010,Oreg_2010} of the topological
superconducting phase emerging in the long chain Kitaev model
\cite{Kitaev_2001}. Experimental evidence for topological superconductivity in
long Kitaev chains has been demonstrated in many labs
\cite{Mourik_2012,Nadj-Perge_2014,Deng_2016,Fornieri_2019,Ren_2019,Wang_2022}.
Nevertheless, these systems still remain hard to control to achieve their
regular technological reproduction avoiding any misinterpretation
\cite{Yu_2021,Frolov_2021} regarding the physical nature of their low-energy
states in favor of non-Majorana states. Indeed, at least in terms of the most
popular physical observables, such as linear conductances, it is not a simple
task to filter out low-energy non-Majorana states resulting in physical
behavior which is difficult to distinguish from what is predicted for MBSs. In
fact, the challenge of observing unique Majorana signatures is one of the
motive forces for many original developments providing deeper insights into
Majorana induced phenomena via a variety of potentially feasible responses of
MBSs. In particular, important and to a large extent independent routes of
exploring MBSs include quantum transport proposals on significantly improved
measurements of various average electric \cite{Liu_2011,Fidkowski_2012,
Prada_2012,Pientka_2012,Lin_2012,Lee_2013,Kundu_2013,Vernek_2014,Ilan_2014,
Lobos_2015,Peng_2015,Sharma_2016,Heck_2016,Das_Sarma_2016,Lutchyn_2017,
Weymann_2017,Campo_Jr_2017,Liu_2017,Huang_2017,Liu_2018,Lai_2019,Tang_2020,
Zhang_2020,Chi_2021,Galambos_2022,Jin_2022,Zou_2023,Huguet_2023,Becerra_2023,
Ziesen_2023,Yao_2023,Taranko_2024,Mondal_2025} and thermoelectric
\cite{Leijnse_2014,Lopez_2014,Khim_2015,Ramos-Andrade_2016,Ricco_2018,
Smirnov_2020a,Wang_2021,He_2021,Giuliano_2022,Buccheri_2022,
Bondyopadhaya_2022,Zou_2022,Wang_2023,Zou_2023a,Chi_2024,Mishra_2024,
Trocha_2025} currents, characterized by corresponding conductances, or, more
advanced ones, on nonequilibrium fluctuations of electric \cite{Liu_2015,
Liu_2015a,Haim_2015,Valentini_2016, Zazunov_2016,Smirnov_2017,Jonckheere_2019,
Bathellier_2019,Smirnov_2019,Manousakis_2020,Smirnov_2022,Feng_2022,Cao_2023,
Smirnov_2024,Barros_2025,Bostroem_2025,Yu_2025} and thermoelectric
\cite{Smirnov_2018,Smirnov_2019a,Smirnov_2023,Smirnov_2025,Mishra_2025}
currents, thermodynamic proposals to measure the unique fractional Majorana
entropy  \cite{Smirnov_2015,Sela_2019,Silva_2020,Smirnov_2021,Smirnov_2021a}
of equilibrium nanosystems hosting MBSs, entanglement measures
\cite{Vimal_2024} of MBSs and aspects of quantum memory \cite{Maroulakos_2025}
in Majorana quantum dot (QD) systems.

An alternative route to explore the wide spectrum of physical phenomena driven
by Majorana degrees of freedom is to resort to the more controllable case of
short, or minimal, Kitaev chains engineered to engage a small number of QDs
into experimentally appealing and more reproducible platforms. The QDs
composing such nanostructures are natural spatial traps for localization of
various Majorana states often termed in these systems as poor man's MBSs
\cite{Leijnse_2012,Fulga_2013}. Similar to MBSs in long Kitaev chains, poor
man's MBSs maintain essentially all the features beneficial for quantum
information processing apart from the strong topological protection. In
particular, these states are non-Abelian anyons characterized by highly
nonlocal spatial distributions. In the simplest case one obtains poor man's
MBSs within a double QD Kitaev chain. Here the QDs interact via a
superconductor placed between them. The superconductor induces between the QDs
both normal tunneling and crossed Andreev reflection processes whose
amplitudes are close to each other resulting in the degeneracy between the
ground states with the opposite fermion parities. In practice these amplitudes
may be tuned, for example, by means of varying the angle between the spins in
the QDs \cite{Leijnse_2012} or the energy of the Andreev bound states in the
superconductor \cite{Liu_2022}. Experimental realizations of short Kitaev
chains and their probes via average electric currents, or the corresponding
conductances, \cite{Dvir_2023,Zatelli_2024,Haaf_2024,Bordin_2025,Haaf_2025}
have been recently started demonstrating a practical feasibility of quantum
transport measurements in these systems which are currently being intensively
developed to involve more QDs. Gradually increasing the number of QDs
establishes a reliable technology for reaching the strong topological
protection of the Majorana states in near future. Meanwhile, already at
present short Kitaev chains provide a convenient platform to build Majorana
qubits \cite{Pino_2024,Pan_2025} for practical applications of Majorana
quasiparticles in quantum computing. Moreover, they allow to probe the
non-Abelian nature of poor man's MBSs via a simpler access to the outcomes of
their fusion and braiding statistics
\cite{Liu_2023,Pandey_2024,Tsintzis_2024,Boross_2024} in comparison with
braiding protocols in long Kitaev chains. In parallel with their applications
in quantum computations poor man's MBSs may play a crucial role in
nonequilibrium phenomena. For example, a recent demonstration of quantum
Mpemba effects \cite{Nava_2024} in a double QD Kitaev chain opens a room to
explore how the poor man's MBSs provide different relaxation paths for diverse
initial nonequilibrium states. Besides the quantum transport spectroscopy, a
microwave response \cite{Dourado_2025} of a short Kitaev chain, implemented
using three QDs, has been introduced as an independent probe of poor man's
MBSs. Focusing on quantum transport, as an established and experimentally
relevant framework, one often deals with nonequilibrium states which result in
electric currents whose average values and deviations from them are
essentially independent of each other. Thus nonequilibrium fluctuations of
electric currents in minimal Kitaev chains, for example nonequilibrium shot
noise, may provide fundamentally different characteristics of poor man's MBSs
and substantially complement average electric currents, or the corresponding
conductances, which, as emphasized above, have been already
measured. Moreover, it also looks plausible that a proper universal
combination of the shot noise and conductance may serve as an experimentally
relevant indicator of the poor man's MBSs both in weakly and strongly
nonequilibrium states.

In this work we analyze fluctuations of electric currents in a minimal Kitaev
chain consisting of two QDs interacting with each other via normal tunneling
and crossed Andreev reflection. To this end we numerically calculate the shot
noise and average electric current in weakly and strongly nonequilibrium
states excited by an external bias voltage applied to one of the QDs coupled
to two normal metallic contacts. To quantitatively characterize the
fluctuation response of the system we use the differential effective charge
$q$ defined as the ratio of the differential shot noise and conductance. This
quantity has universal units of the elementary charge $e$ and offers practical
advantages in comparison with separate measurements of the differential shot
noise and conductance both in weakly and strongly nonequilibrium minimal
Kitaev chains.

Specifically, we demonstrate that the Majorana sweet spot region, that is a
certain parameter region where the poor man's MBSs largely drive the
fluctuation behavior of the system, is characterized by two fractional values
of the differential effective charge, $q=e/2$ and $q=3e/2$. The fractional
value $q=e/2$ is observed only within extremely narrow vicinities of specific
points of the Majorana sweet spot region which are the points where the
difference between the amplitudes of the normal tunneling and crossed Andreev
reflection becomes equal to the bias voltage. At the same time, within the
main body of the Majorana sweet spot region the differential effective charge
fractionalizes to $q=3e/2$. Obviously, since the domains with $q=e/2$ are
extremely narrow, it is highly probable that in possible experiments one will
detect the fractional value $q=3e/2$ and not $q=e/2$. Thus the fractional
value $q=3e/2$ may be used as a reliable indicator of the Majorana sweet spot
region both in weakly and strongly nonequilibrium systems. It has a practical
advantage over separate measurements of the differential shot noise and
conductance in situations where each of these quantities is not sufficiently
informative. For example, this may happen within the Majorana sweet spot
region in strongly nonequilibrium minimal Kitaev chains, when the differential
shot noise and conductance have a featureless, non-resonant behavior. Also in
weakly nonequilibrium minimal Kitaev chains the differential shot noise and
conductance may be noticeably below their universal unitary limits. Indeed, it
is reasonable to assume that in experiments it may be possible to achieve only
an approximate equality between the amplitudes of the normal tunneling and
crossed Andreev reflection but not the exact equality. As a result, the system
turns out to be within the Majorana sweet spot region but not at its
center. As a consequence, the differential shot noise and conductance do not
reach their universal Majorana values. In contrast, in all the above
situations the differential effective charge fractionalizes to $q=3e/2$ and
reliably reveals that exactly the poor man's MBSs vastly contribute to the
fluctuation behavior of the system.

The paper is organized as follows. In Section \ref{Hamiltonian} we specify the
total Hamiltonian of the system which includes the double QD Kitaev chain and
normal metallic contacts coupled to one of the QDs. The basic aspects of the
Keldysh field integral used to calculate the shot noise and average electric
current are provided in Section \ref{Keldysh}. Numerical results obtained for
the ratio of the differential shot noise and conductance, that is for the
differential effective charge of the minimal Kitaev chain, are presented in
Section \ref{Results} in the regime of both low and high bias
voltages. Finally, with Section \ref{Conclusion} we briefly summarize the
paper and draw conclusions.
\section{Hamiltonian of a minimal Kitaev chain composed of two QDs}\label{Hamiltonian}
\begin{figure}
  \includegraphics[width=8.0 cm]{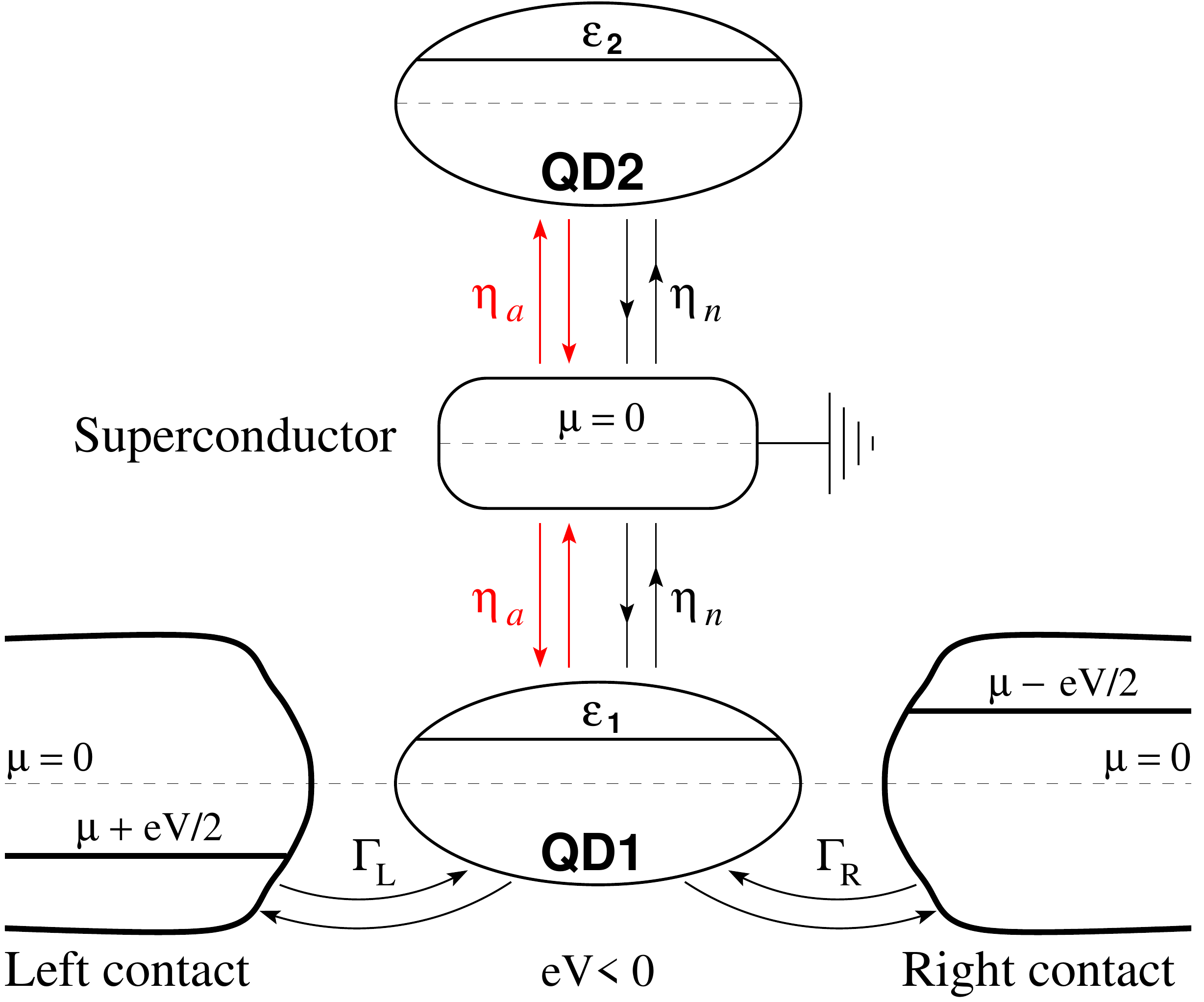}
  \caption{\label{figure_1} A schematic representation of the system composed
    of two QDs, QD1 and QD2, grounded superconductor inbetween and two normal
    metallic contacts. Each QD is characterized by one non-degenerate
    single-particle energy level, $\varepsilon_1$ and $\varepsilon_2$,
    localized, respectively, in QD1 and QD2. The superconductor governs
    both the normal tunneling and crossed Andreev reflection between the
    QDs. The corresponding processes occur with the amplitudes $|\eta_n|$ and
    $|\eta_a|$. Both contacts are coupled to QD1 via normal tunneling of
    strength $\Gamma_{L,R}$. The difference in their chemical potentials is
    the direct source for nonequilibrium states in this system and the degree
    of nonequilibrium is controlled by the bias voltage $V$. The Hamiltonian
    of this composite nanostructure is specified in
    Eqs. (\ref{Ham})-(\ref{Symm_coupling}).}
\end{figure}

The Hamiltonian of the system,
\begin{equation}
  \hat{H}=\hat{H}_\text{MKC}+\hat{H}_\text{C}+\hat{H}_\text{MKC-C},
\label{Ham}
\end{equation}
is the sum of the Hamiltonians describing the minimal Kitaev chain, two normal
metallic contacts and tunneling between the minimal Kitaev chain and
contacts. The whole system is illustrated in Fig. \ref{figure_1}.

The Hamiltonian of the double QD Kitaev chain \cite{Leijnse_2012} has the
following form:
\begin{equation}
  \hat{H}_\text{MKC}=\hat{H}_\text{D1}+\hat{H}_\text{D2}+\hat{H}_\text{D1-D2}.
  \label{Ham_MKC}
\end{equation}
Here $\hat{H}_\text{D1}$ and $\hat{H}_\text{D2}$ are the Hamiltonians of the
QDs, denoted as QD1 and QD2:
\begin{equation}
  \hat{H}_\text{D1}=\varepsilon_1d_1^\dagger d_1,\quad
  \hat{H}_\text{D2}=\varepsilon_2d_2^\dagger d_2,
  \label{Ham_D1_Ham_D2}
\end{equation}
where $\varepsilon_1$ and $\varepsilon_2$ are the energies of the
non-degenerate single-particle states localized in QD1 and QD2. Both
$\varepsilon_1$ and $\varepsilon_2$ may be tuned by appropriate gate voltages
applied to the corresponding QDs. The Hamiltonian $\hat{H}_\text{D1-D2}$
describes the coupling between QD1 and QD2:
\begin{equation}
  \hat{H}_\text{D1-D2}=\hat{H}_\text{NT}+\hat{H}_\text{CAR}.
  \label{Ham_D1-D2}
\end{equation}
This interdot coupling is implemented by a superconductor placed between the
QDs and accounts for the normal tunneling between QD1 and QD2,
\begin{equation}
  \hat{H}_\text{NT}=\eta_n^\star d_1^\dagger d_2+\text{H.c.},
  \label{Ham_NT}
\end{equation}
as well as for the crossed Andreev reflection between them,
\begin{equation}
  \hat{H}_\text{CAR}=\eta_a^\star d_1^\dagger d_2^\dagger+\text{H.c.},
  \label{Ham_CAR}
\end{equation}
where $\eta_n$ and $\eta_a$ are the matrix elements of the normal tunneling
and crossed Andreev reflection with the corresponding amplitudes $|\eta_n|$
and $|\eta_a|$ which may be both tuned in experiments
\cite{Leijnse_2012,Dvir_2023}.

The two normal metallic contacts, labeled as left ($L$) and right ($R$), are
modeled by the following Hamiltonian:
\begin{equation}
  \hat{H}_\text{C}=\sum_{l=L,R}\sum_k\epsilon_kc_{lk}^\dagger c_{lk}.
  \label{Ham_Cont}
\end{equation}
The contacts have the same energy spectrum $\epsilon_k$ which is continuous
and results in the density of states $\nu(\epsilon)$ whose energy dependence
is often neglected,
\begin{equation}
  \nu(\epsilon)\approx\frac{1}{2}\nu_\text{C},
  \label{DOS_cont}
\end{equation}
in various quantum transport experiments restricted to certain finite energy
domains well described by this approximation. The equilibrium states of the
two contacts may be different, as specified by their Fermi-Dirac
distributions,
\begin{equation}
  n_{L,R}(\epsilon)=\frac{1}{\exp(\frac{\epsilon-\mu_{L,R}}{k_\text{B}T})+1}
  \label{FD_distrib}
\end{equation}
with temperature $T$ and in general different chemical potentials,
\begin{equation}
  \mu_{L,R}=\mu\pm eV/2,
  \label{Chem_pot}
\end{equation}
where $V$ is the bias voltage which, for definiteness, is chosen to be
negative, $eV<0$ (only in Fig. \ref{figure_4}(b) both $eV<0$ and $eV>0$ are
used to explicitly show the resonance of the differential conductance).

The tunneling between the minimal Kitaev chain and contacts is specified by
the configuration within which the two normal metallic contacts interact with
one of the QDs, namely QD1. The corresponding tunneling Hamiltonian is
\begin{equation}
  \hat{H}_\text{MKC-C}=\sum_{l=L,R}\mathcal{T}_l\sum_kc_{lk}^\dagger d_1+\text{H.c.}
  \label{Ham_MKC-C}
\end{equation}
Here we have assumed that $\mathcal{T}_{lk}\approx\mathcal{T}_l$. The
interaction between the minimal Kitaev chain and contacts is characterized by
the energies
\begin{equation}
  \Gamma_{L,R}=\pi\nu_\text{C}|\mathcal{T}_{L,R}|^2.
  \label{Gamma_L_R}
\end{equation}
The total energy of the tunneling coupling between the minimal Kitaev chain
and contacts is
\begin{equation}
  \Gamma=\Gamma_L+\Gamma_R.
  \label{Gamma}
\end{equation}
In this work, for simplicity, we investigate the system with the symmetric
coupling of the minimal Kitaev chain to the left and right contacts,
\begin{equation}
  \Gamma_L=\Gamma_R.
  \label{Symm_coupling}
\end{equation}

In our numerical calculations we mainly focus on the regime where the energy
of the interdot coupling dominates over the other energy scales in the
system:
\begin{equation}
  \sqrt{|\eta_n|^2+|\eta_a|^2}>\max\{|\varepsilon_1|,|\varepsilon_2|,k_\text{B}T,\Gamma,|eV|\}.
  \label{Univ_Majorana_reg}
\end{equation}
Within this regime and for proper ratios between $|\eta_n|$ and $|\eta_a|$ the
poor man's MBSs may emerge and provide major contributions to quantum
transport observables. As will be shown in Section \ref{Results}, these
Majorana contributions turn out to be universal, that is they become
independent of the corresponding gate voltage. Thus we will refer to the
regime specified in Eq. (\ref{Univ_Majorana_reg}) as to the universal Majorana
regime even though Eq. (\ref{Univ_Majorana_reg}) also admits (see Section
\ref{Results}) non-Majorana behavior if $|\eta_n|$ and $|\eta_a|$ are chosen
far outside the Majorana sweet spot region.
\section{Keldysh action for the double quantum dot Kitaev chain and
  nonequilibrium shot noise}\label{Keldysh}
To comprehensively explore various nonequilibrium states emerging in our
system in response to the time independent bias voltage $V$ as well as to
straightforwardly obtain the average electric current and shot noise, we
resort to the nonequilibrium Keldysh technique \cite{Keldysh_1965,Landau_X}
within its field integral formulation \cite{Altland_2023}. Its basic elements
for nonequilibrium Majorana systems, have been presented previously (see,
{\it e.g}, Ref. \cite{Smirnov_2018}). With this in mind we run shortly through
the key aspects of the Keldysh field integral and dwell only on those details
in the Keldysh action which are relevant for minimal Kitaev chains.

Since the minimal Kitaev chain and contacts are fermionic systems, the Keldysh
field integral is naturally written in terms of the Grassmann fields,
$(\psi_1,\bar{\psi}_1)$, $(\psi_2,\bar{\psi}_2)$ and
$(\phi_{lk},\bar{\phi}_{lk})$, where the bars stand for the Grassmann
conjugation (G.c.). These Grassmann pairs correspond to the pairs of the
annihilation and creation operators $(d_1,d_1^\dagger)$, $(d_2,d_2^\dagger)$
and $(c_{lk},c_{lk}^\dagger)$ of QD1, QD2 and the contacts. The nonequilibrium
dynamics evolves along the Keldysh contour which is a closed time contour. It
has a forward and backward branch numbered with the discrete index
$q=\pm$. The evolution along these branches is specified via the real time
$t$. This conveniently splits the Keldysh contour and allows to obtain various
correlation functions in real time. It is achieved by means of the Keldysh
generating functional which may be expressed as a Grassmann field integral,
\begin{equation}
  \mathcal{Z}[\mathcal{J}_{lq}(t)]=\int\mathcal{D}[\bar{\mathcal{X}}_q(t),\mathcal{X}_q(t)]\exp\biggl\{\frac{i}{\hbar}\mathcal{S}_\text{K}[\mathcal{J}_{lq}(t)]\biggr\},
  \label{KGF}
\end{equation}
where
\begin{equation}
  \begin{split}
    &\mathcal{X}_q(t)=[\psi_{1q}(t),\psi_{2q}(t),\phi_{lkq}(t)],\\
    &\bar{\mathcal{X}}_q(t)=[\bar{\psi}_{1q}(t),\bar{\psi}_{2q}(t),\bar{\phi}_{lkq}(t)],
  \end{split}
  \label{Grassmann_fields_pm}
\end{equation}
\begin{equation}
  \begin{split}
    &\mathcal{S}_\text{K}[\mathcal{J}_{lq}(t)]=\sum_{i=1}^2S_{\text{D}i}[\bar{\psi}_{iq}(t),\psi_{iq}(t)]\\
    &+S_\text{D1-D2}[\bar{\psi}_{1q}(t),\bar{\psi}_{2q}(t);\psi_{1q}(t),\psi_{2q}(t)]\\
    &+S_\text{C}[\bar{\phi}_{lkq}(t),\phi_{lkq}(t)]\\
    &+S_\text{MKC-C}[\bar{\psi}_{1q}(t),\bar{\phi}_{lkq}(t);\psi_{1q}(t),\phi_{lkq}(t)]\\
    &+S_\text{SRC}[\mathcal{J}_{lq}(t)].
  \end{split}
  \label{Keldysh_action}
\end{equation}
Here $S_{\text{D}i}$, $i=1,2$, and $S_\text{D1-D2}$ are the actions of QD1,
QD2 and their coupling whereas the actions $S_\text{C}$ and $S_\text{MKC-C}$
describe the contacts and their coupling to the minimal Kitaev chain. The last
term $S_\text{SRC}$ is the source action used to generate various observables
via proper derivatives over the sources $\mathcal{J}_{lq}(t)$ taken at
$\mathcal{J}_{lq}(t)=0$ and assuming the normalization
$S_\text{SRC}[\mathcal{J}_{lq}(t)=0]=1$. Further, the Keldysh rotation
\cite{Altland_2023} of the Grassmann fields $\mathcal{X}_q(t)$ and
$\bar{\mathcal{X}}_q(t)$ in Eq. (\ref{Grassmann_fields_pm}) is applied to mix
the froward and backward branches of the Keldysh contour to formulate the
theory in terms of the retarded, advanced and Keldysh correlation
functions. This leads to the conventional $2\times 2$ upper triangular matrix
structure \cite{Altland_2023} of the actions for QD1, QD2 and the contacts.
The action for the coupling between QD1 and QD2,
\begin{equation}
  \begin{split}
    &S_\text{D1-D2}[\bar{\psi}_{1q}(t),\bar{\psi}_{2q}(t);\psi_{1q}(t),\psi_{2q}(t)]\\
    &=-\!\!\int_{-\infty}^\infty \!\!\!\!\!dt\sum_qq\{[\eta_n^\star\bar{\psi}_{1q}(t)\psi_{2q}(t)+\eta_a^\star\bar{\psi}_{1q}(t)\bar{\psi}_{2q}(t)]\\
    &\quad\quad\quad\quad\quad\quad\quad+\text{G.c.}\},
  \end{split}
  \label{Action_D1-D2}
\end{equation}
as well as between the minimal Kitaev chain and contacts,
\begin{equation}
  \begin{split}
    &S_\text{MKC-C}[\bar{\psi}_{1q}(t),\bar{\phi}_{lkq}(t);\psi_{1q}(t),\phi_{lkq}(t)]\\
    &=-\int_{-\infty}^\infty dt\sum_{l=L,R}\sum_{k,q}q[\mathcal{T}_l\bar{\phi}_{lkq}(t)\psi_{1q}(t)+\text{G.c.}],
  \end{split}
  \label{Action_MKC-C}
\end{equation}
are also transformed by the Keldysh rotation.

Since our goal is to calculate the average electric current and shot noise, we
specify the source action using the operator of the electric current measured
in a given contact $l=L,R$,
\begin{equation}
  \hat{I}_l=\frac{ie}{\hbar}\sum_k(\mathcal{T}_lc_{lk}^\dagger d_1-\text{H.c.}),
  \label{Current_op}
\end{equation}
via coupling the corresponding Grassmann expression,
\begin{equation}
  I_{lq}(t)=\frac{ie}{\hbar}\sum_k[\mathcal{T}_l\bar{\phi}_{lkq}(t)\psi_{1q}(t)-\text{G.c.}],
  \label{Current_Gr}
\end{equation}
to the sources,
\begin{equation}
  S_\text{SRC}[\mathcal{J}_{lq}(t)]=-\int_{-\infty}^\infty dt\sum_{l=L,R}\sum_q\mathcal{J}_{lq}(t)I_{lq}(t).
  \label{Source_action}
\end{equation}
Having defined such a source action, one may derive various current-current
correlation functions. Of particular interest to us are the average electric
current measured in contact $l'$,
\begin{equation}
  I_{l'}(V)=\langle I_{l'q'}(t')\rangle_0=i\hbar\frac{\delta\mathcal{Z}[\mathcal{J}_{lq}(t)]}{\delta\mathcal{J}_{l'q'}(t')}\biggr|_{\mathcal{J}_{lq}(t)=0},
  \label{Av_current}
\end{equation}
and also the correlation between the two values of the current measured in
contacts $l_1$, $l_2$ at two arbitrary real times $t_1$, $t_2$ taken on the
Keldysh branches, $q_1$, $q_2$,
\begin{equation}
  \begin{split}
    &\langle I_{l_1q_1}(t_1)I_{l_2q_2}(t_2)\rangle_0\\
    &=(i\hbar)^2\frac{\delta^2\mathcal{Z}[\mathcal{J}_{lq}(t)]}{\delta\mathcal{J}_{l_1q_1}(t_1)\delta\mathcal{J}_{l_2q_2}(t_2)}\biggr|_{\mathcal{J}_{lq}(t)=0},
  \end{split}
  \label{Current-Current_Corr}
\end{equation}
where
\begin{equation}
  \begin{split}
    &\langle I_{l_1q_1}(t_1)\cdots I_{l_nq_n}(t_n)\rangle_0\\
    &=\int\mathcal{D}[\bar{\mathcal{X}}_q(t),\mathcal{X}_q(t)]e^{\frac{i}{\hbar}\mathcal{S}_\text{K}^{(0)}}I_{l_1q_1}(t_1)\cdots I_{l_nq_n}(t_n),
  \end{split}
  \label{Average_with_SK0}
\end{equation}
\begin{equation}
  \mathcal{S}_\text{K}^{(0)}=\mathcal{S}_\text{K}[\mathcal{J}_{lq}(t)=0].
  \label{SK0}
\end{equation}
The calculations of various averages $\langle\cdots\rangle_0$ may be done via
a suitable form of the Wick's theorem as has been presented, for example in
Ref. \cite{Smirnov_2018}, which provides further technicalities.

In this work we focus on the electric current measured in the left contact
where we calculate its average value, $I(V)=I_L(V)$, and shot noise,
$S(V)$. To obtain the shot noise $S(V)$ we define the greater current-current
correlation function which measures fluctuations of the electric current
around its average value:
\begin{equation}
  S^>(t-t',V)=\langle\delta I_{L-}(t)\delta I_{L+}(t')\rangle_0,
  \label{Greater_noise}
\end{equation}
\begin{equation}
  \delta I_{lq}(t)=I_{lq}(t)-I_l(V),\quad l=L,R.
  \label{Current_fluct_Gr}
\end{equation}
Then its Fourier transform,
\begin{equation}
  S^>(\omega,V)=\int_{-\infty}^\infty dt e^{i\omega t}S^>(t,V),
  \label{Greater_noise_fr}
\end{equation}
taken at zero frequency, provides the shot noise:
\begin{equation}
  S(V)=S^>(\omega=0,V).
  \label{Shot_noise}
\end{equation}

An important quantity relevant to characterize fluctuations of electric
currents is the ratio of the derivatives of the shot noise and average
electric current with respect to the bias voltage. This Fano-like quantity is
measured in units of the elementary charge $e$ and often allows to clearly
reveal important universal properties of the fluctuation behavior in various
systems, in particular in those supporting anyon excitations. Indeed, the Fano
factor turns out to be extremely fruitful
\cite{de-Picciotto_1997,Saminadayar_1997} for analyzing physical phenomena
driven by the fractional quantum Hall effect where the relevant degrees of
freedom are anyon excitations know as Laughlin quasiparticles. Thus it is
quite natural to explore this ratio in minimal Kitaev chains able to support
poor man's MBSs which are also anyon excitations. To this end, we define the
differential effective charge as the ratio of the differential shot noise and
conductance,
\begin{equation}
  q=\frac{\partial_V S(V)}{\partial_V I(V)},
  \label{Diff_eff_charge}
\end{equation}
which is numerically analyzed in the next section for both weakly and strongly
nonequilibrium states quantified by the magnitude of the bias voltage $V$ or
the corresponding energy scale $|eV|$ in comparison with the other energy
scales of the minimal Kitaev chain.
\section{Numerical results on nonequilibrium fluctuation behavior of the
  minimal Kitaev chain}\label{Results}
In this section we present numerical results for the differential effective
charge $q$ defined in the previous section, Eq. (\ref{Diff_eff_charge}). To
obtain the results we have first calculated $I(V)$ and $S(V)$ via numerical
integrations and after that applied finite differences to numerically
calculate the derivatives $\partial I(V)/\partial V$ and
$\partial S(V)/\partial V$ whose ratio provides the differential effective
charge $q$ as a function of various parameters of the system.

\begin{figure}
  \includegraphics[width=8.0 cm]{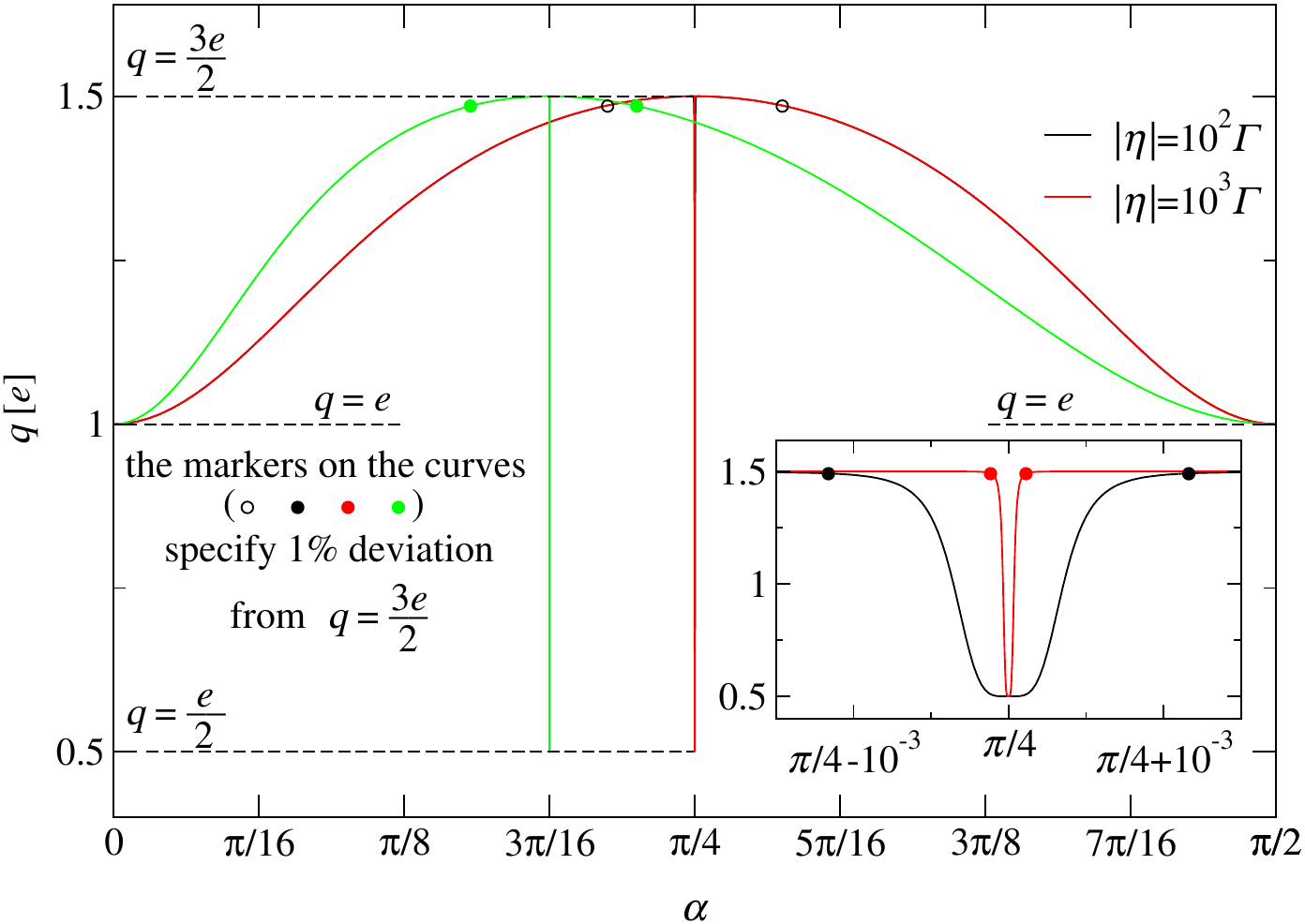}
  \caption{\label{figure_2} The differential effective charge,
    $q\equiv\partial_VS/\partial_VI$, is shown in universal units of the
    elementary charge $e$ as a function of the ratio between the amplitudes
    $|\eta_n|$ and $|\eta_a|$ for low bias voltages,
    $|eV|\ll\Gamma$. Specifically, $|eV|/\Gamma=10^{-2}$. For the black and
    red curves $|\eta_n|=|\eta|\cos\alpha$, $|\eta_a|=|\eta|\sin\alpha$ with
    $|\eta|/\Gamma=10^2$ (black curve) and $|\eta|/\Gamma=10^3$ (red
    curve). For the green curve $|\eta_n|=|\eta_1|\cos\alpha$,
    $|\eta_a|=|\eta_2|\sin\alpha$ with $|\eta_1|/\Gamma=10^3$,
    $|\eta_2|=|\eta_1|\cot(3\pi/16)$. The values of the other parameters:
    $k_\text{B}T/\Gamma=10^{-7}$, $\varepsilon_1/\Gamma=10$,
    $\varepsilon_2/\Gamma=10^{-9}$. The inset illustrates in more detail a
    very narrow vicinity of the point $\alpha=\pi/4$.}
\end{figure}

In Fig. \ref{figure_2} we show the differential effective charge $q$ as a
function of the ratio $|\eta_n|/|\eta_a|$ for small bias voltages,
$|eV|\ll\Gamma$. To conveniently parameterize the two real amplitudes
$|\eta_n|$ and $|\eta_a|$ we introduce two real variables $|\eta|$ and
$\alpha$,
\begin{equation}
  |\eta_n|=|\eta|\cos\alpha,\quad|\eta_a|=|\eta|\sin\alpha,
  \label{Param_eta_n_eta_a}
\end{equation}
where the angular variable, $0\leqslant\alpha\leqslant\pi/2$, parameterizes
the ratio $|\eta_n|/|\eta_a|$. The point where $|\eta_n|=|\eta_a|$ is
specified by $\alpha=\pi/4$. As we can see, for small bias voltages the sweet
spot region, that is the neighborhood of the point $|\eta_n|=|\eta_a|$, or,
equivalently, the neighborhood of the point $\alpha=\pi/4$, is characterized
by two fractional values of the differential effective charge, $q=e/2$ and
$q=3e/2$. Specifically, exactly at the point $\alpha=\pi/4$ the differential
effective charge takes the fractional value $q=e/2$. As demonstrated in the
inset, this fractional value is observed in an extremely narrow vicinity of
the point $\alpha=\pi/4$. Numerically we find that the differential effective
charge quickly grows from the fractional value $q=e/2$ to the fractional value
$q=3e/2$ when one moves slightly away from the point $\alpha=\pi/4$. For
example, 1\% deviation from the fractional value $q=3e/2$ (see the black and
red dots on the curves in the inset) is reached for the absolute difference
$|\eta_n|-|\eta_a|\approx\pm 0.143\Gamma$ independently of $|\eta|$, that is
both for the black and red curves. In terms of the relative difference, it is
reached for
$(|\eta_n|-|\eta_a|)/\text{max}\{|\eta_n|,|\eta_a|\}\approx\pm 2.0\times 10^{-3}$
(or $0.2\%$) if $|\eta|/\Gamma=10^2$ (black curve) and for
$(|\eta_n|-|\eta_a|)/\text{max}\{|\eta_n|,|\eta_a|\}\approx\pm 2.0\times 10^{-4}$
(or $0.02\%$) if $|\eta|/\Gamma=10^3$ (red curve). Outside this very narrow
vicinity of $\alpha=\pi/4$ the differential effective charge takes the
fractional value $q=3e/2$ which is observed in the main body of the sweet
spot region. At this point it is important to note that in experiments usually
one deals with systems where $|\eta_n|$ and $|\eta_a|$ are only close to each
other, $|\eta_n|\approx|\eta_a|$, but not exactly equal. This circumstance and
the fact that the region with $q=e/2$ is extremely narrow imply that it is
most probable that in experiments one will observe the fractional value
$q=3e/2$ and not $q=e/2$. Thus the fractional value $q=3e/2$ may serve as an
indicator of the Majorana sweet spot region located around $\alpha=\pi/4$,
where the poor man's MBSs largely drive fluctuations of the electric
current. In particular, it can be used to estimate the size of the Majorana
sweet spot region. Indeed, as we can see in the main plot of
Fig. \ref{figure_2}, moving away from the Majorana sweet spot region reduces
the differential effective charge well below $3e/2$. If we, for example,
define the boundary of the Majorana sweet spot region as the set of points at
which the differential effective charge is 1\% less than $q=3e/2$ (black
circles in the main plot), then, in terms of the absolute difference, the
boundary is located at $(|\eta_n|-|\eta_a|)\approx\pm 17.2\Gamma$ for
$|\eta|/\Gamma=10^2$ and at $(|\eta_n|-|\eta_a|)\approx\pm 172\Gamma$ for
$|\eta|/\Gamma=10^3$. Since outside the very narrow vicinity, shown in the
inset, both the black and red curves almost coincide (including the black
circles in the main plot), we see that in terms of the relative difference the
boundary of the Majorana sweet spot region does not depend on $|\eta|$ and is
characterized by
$(|\eta_n|-|\eta_a|)/\text{max}\{|\eta_n|,|\eta_a|\}\approx\pm 0.22$ (or
$22\%$). When one runs away from the Majorana sweet spot region, the
differential effective charge monotonously decreases and reaches the trivial
integer value $q=e$ at the edge points, $\alpha=0$ ($|\eta_n|=|\eta|$,
$|\eta_a|=0$) and $\alpha=\pi/2$ ($|\eta_n|=0$, $|\eta_a|=|\eta|$), where it
does not depend on the magnitude of $|\eta|$. In other words, when one of the
amplitudes vanishes, $|\eta_a|=0$ or $|\eta_n|=0$, the differential effective
charge does not depend on the other amplitude which remains finite, that is,
respectively, on $|\eta_n|$ or $|\eta_a|$, and retains its trivial integer
value $q=e$. For practical purposes it may also be useful to consider a more
general parameterization,
\begin{equation}
  |\eta_n|=|\eta_1|\cos\alpha,\quad|\eta_a|=|\eta_2|\sin\alpha,
  \label{Param_eta_n_eta_a_gen}
\end{equation}
where the angular variable, $0\leqslant\alpha\leqslant\pi/2$, still
parameterizes the ratio $|\eta_n|/|\eta_a|$ but, in contrast to the
parameterization in Eq. (\ref{Param_eta_n_eta_a}), now we assume that
$|\eta_1|\neq|\eta_2|$. It turns out that in this case one also observes all
the key aspects, discussed above for the particular choice
$|\eta_1|=|\eta_2|=|\eta|$. The only difference is that now the differential
effective charge $q$ as a function of $\alpha$ becomes asymmetric with respect
to the central point $\alpha=\pi/4$. For example, the green curve in
Fig. \ref{figure_2} shows the differential effective charge obtained for
$|\eta_2|=|\eta_1|\cot(3\pi/16)$. With this parameterization we have
$|\eta_n|=|\eta_a|$ for $\alpha=3\pi/16$. Thus the Majorana sweet spot region
is now located around the point $\alpha=3\pi/16$ as demonstrated by the green
curve which is obviously asymmetric with respect to $\alpha=\pi/4$. Except for
the shift of the Majorana sweet spot region and the asymmetry of the
differential effective charge, the basic features revealed previously for the
symmetric parameterization, $|\eta_1|=|\eta_2|=|\eta|$, are all preserved. In
particular, when $|\eta_1|\neq|\eta_2|$, the Majorana sweet spot region is
still characterized by the two fractional values $q=e/2$ and $q=3e/2$; the
fractional value $q=e/2$ is only observed in a very narrow vicinity of the
point $|\eta_n|=|\eta_a|$ whereas the main body of the sweet spot is
characterized by the fractional value $q=3e/2$; the size of the Majorana sweet
spot region (green dots on the green curve) is almost the same; at the edge
points, $\alpha=0,\pi/2$, the differential effective charge is still equal to
the trivial integer value $q=e$.
\begin{figure}
  \includegraphics[width=8.0 cm]{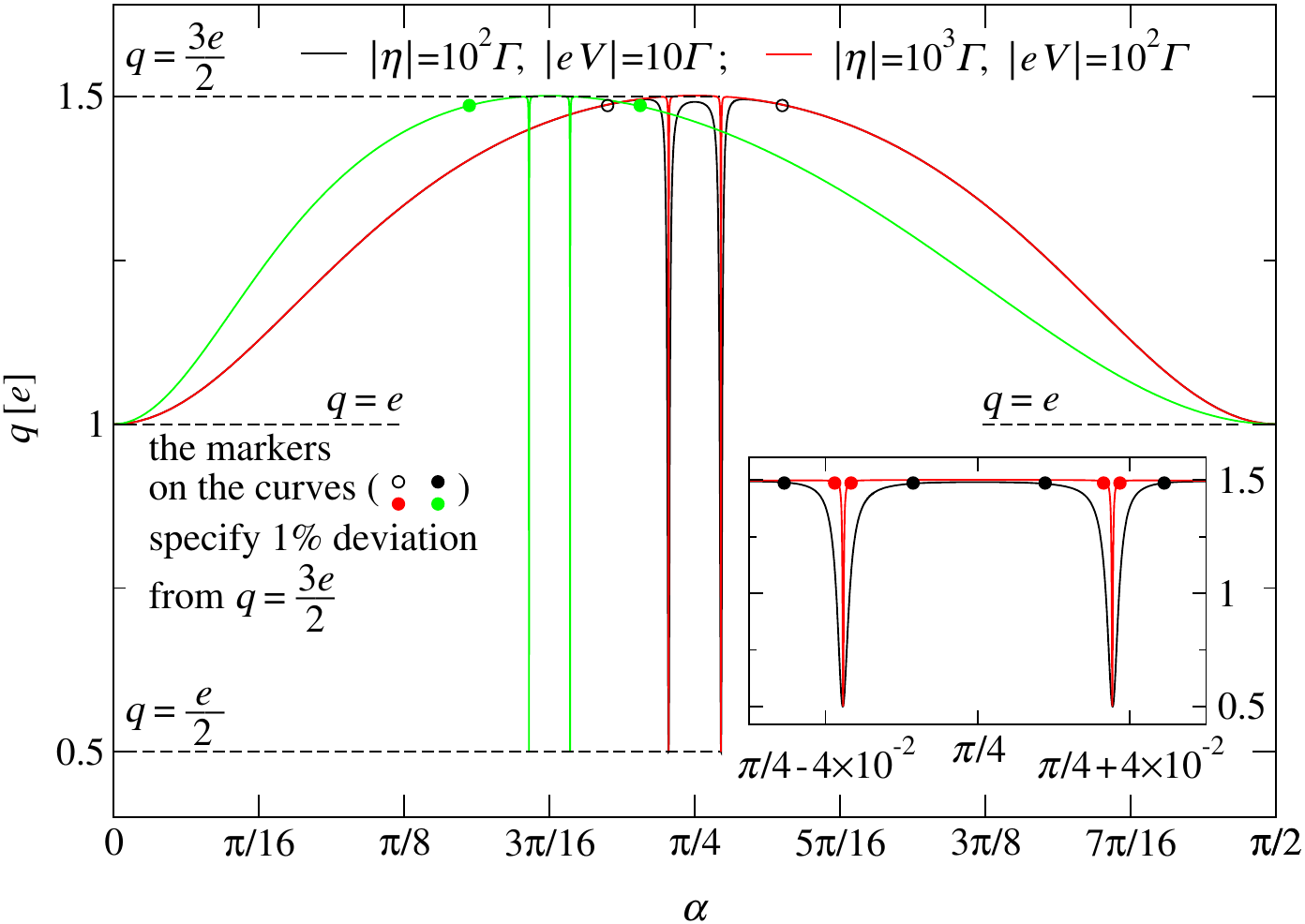}
  \caption{\label{figure_3} The differential effective charge,
    $q\equiv\partial_VS/\partial_VI$, is shown in universal units of the
    elementary charge $e$ as a function of the ratio between the amplitudes
    $|\eta_n|$ and $|\eta_a|$ for high bias voltages,
    $|eV|\gg\Gamma$. Specifically, $|eV|/\Gamma=10$ for the black curve and
    $|eV|/\Gamma=10^2$ for the red and green curves. For the black and red
    curves $|\eta_n|=|\eta|\cos\alpha$, $|\eta_a|=|\eta|\sin\alpha$ with
    $|\eta|/\Gamma=10^2$ (black curve) and $|\eta|/\Gamma=10^3$ (red
    curve). For the green curve $|\eta_n|=|\eta_1|\cos\alpha$,
    $|\eta_a|=|\eta_2|\sin\alpha$ with $|\eta_1|/\Gamma=10^3$,
    $|\eta_2|=|\eta_1|\cot(3\pi/16)$. The values of the other parameters are
    the same as in Fig. \ref{figure_2}. The inset zooms in relevant details in
    a vicinity of the point $\alpha=\pi/4$.}
\end{figure}

Now, let us consider how the fluctuation behavior discussed above for weakly
nonequilibrium states ($|eV|\ll\Gamma$) transforms when the minimal Kitaev
chain is brought into strong nonequilibrium by large bias voltages,
$\sqrt{|\eta_n|^2+|\eta_a|^2}>|eV|\gg\Gamma$. The behavior of the differential
effective charge $q$ in this case is illustrated in Fig. \ref{figure_3} as a
function of the ratio $|\eta_n|/|\eta_a|$. Here for the black and red curves
we use the parameterization introduced in Eq. (\ref{Param_eta_n_eta_a}). As we
can see, in contrast to the case of low bias voltages, where the very narrow
region with $q=e/2$ forms around the point $\alpha=\pi/4$
($|\eta_n|=|\eta_a|$), such a region does not appear in the case of large bias
voltages. Nevertheless, the differential effective charge remains fractional
at the point $|\eta_n|=|\eta_a|$ but now it fractionalizes to $q=3e/2$. At the
same time, there develop two very narrow regions with the fractional value
$q=e/2$ located, as we find numerically, at the points
\begin{equation}
  |\eta_n|-|\eta_a|=\pm\frac{|eV|}{2},
  \label{Loc_of_e_half_high_V}
\end{equation}
in full agreement with the analytical energies presented in
Ref. \cite{Leijnse_2012} for the double QD Kitaev chain. We note that since
the ratio $|eV|/|\eta|$ is the same for both the red and black curves, their
regions with $q=e/2$ are located around the same values of $\alpha$, in full
accordance with Eq. (\ref{Loc_of_e_half_high_V}). Within these two very narrow
regions the differential effective charge quickly grows from $q=e/2$ to
$q=3e/2$ for relatively small shifts from their centers. The black and red
dots on the corresponding curves in the inset show where the deviation of the
differential effective charge from the fractional value $q=3e/2$ already
reaches 1\%. Since these two regions are very narrow, they do not perturb the
rest parts of the curves which remain almost the same as in the low bias
regime. Indeed, comparing the main plots in Figs. \ref{figure_2} and
\ref{figure_3}, we see that, except for the narrow regions with $q=e/2$, the
black and red curves in Fig. \ref{figure_3} coincide with those in
Fig. \ref{figure_2}. In particular, the boundaries of the Majorana sweet spot
region remain the same as indicated by the black circles in the main plots of
Figs. \ref{figure_2} and \ref{figure_3} (1\% deviation from $q=3e/2$). The
relevant characteristics of the strongly nonequilibrium behavior analyzed
above with the parameterization in Eq. (\ref{Param_eta_n_eta_a}) are also
observed using the general parameterization introduced in
Eq. (\ref{Param_eta_n_eta_a_gen}) as demonstrated by the green curve in
Fig. \ref{figure_3}. In particular, in contrast to the case $|eV|\ll\Gamma$,
the very narrow region with $q=e/2$ does not appear around the point
$|\eta_n|=|\eta_a|$ or $\alpha=3\pi/16$ for the choice of $|\eta_1|$ and
$|\eta_2|$ used in Fig. \ref{figure_3}. The effective charge at
$|\eta_n|=|\eta_a|$ still remains fractional but with the value $q=3e/2$. The
two very narrow regions with $q=e/2$ also form but now, in accordance with
Eq. (\ref{Loc_of_e_half_high_V}), they accompany and surround the point
$\alpha=3\pi/16$ and not $\alpha=\pi/4$. As can be seen from comparison of the
green curves in Figs. \ref{figure_2} and \ref{figure_3}, they coincide except
for the regions with $q=e/2$. Thus also with the general parameterization in
Eq. (\ref{Param_eta_n_eta_a_gen}) the Majorana sweet spot region turns out to
be stable against large bias voltages as confirmed by its unaltered boundaries
labeled in Figs. \ref{figure_2} and \ref{figure_3} by the green dots at which
the differential effective charge becomes 1\% smaller than the fractional
value $q=3e/2$.
\begin{figure}
  \includegraphics[width=8.0 cm]{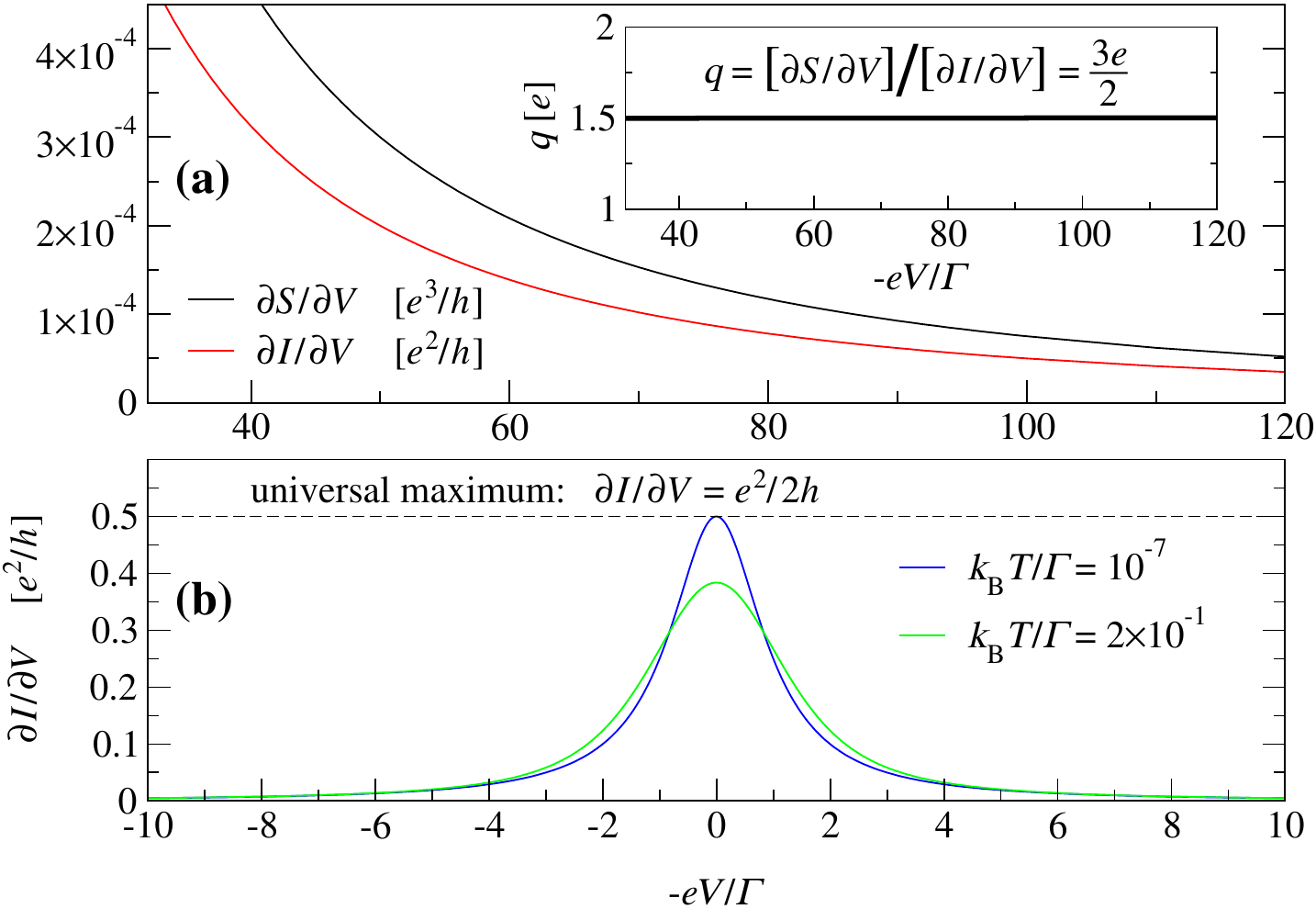}
  \caption{\label{figure_4} Panel ({\bf a}). $\partial_VS$ and $\partial_VI$
    as functions of the bias voltage,
    $\Gamma\ll|eV|<\sqrt{|\eta_n|^2+|\eta_a|^2}$. Here
    $|\eta_n|=|\eta|\cos\alpha$, $|\eta_a|=|\eta|\sin\alpha$ with
    $|\eta|/\Gamma=10^3$ and $\alpha=\pi/4$ (or $|\eta_n|=|\eta_a|$). These
    high voltage tails do not depend on the temperature: the
    black ($\partial_VS$) and red ($\partial_VI$) curves remain unchanged when
    the temperature is drastically increased from $k_\text{B}T/\Gamma=10^{-7}$
    to $k_\text{B}T/\Gamma=2\times 10^{-1}$. The values of the other
    parameters are the same as in Fig. \ref{figure_2}. Inset:
    $q\equiv\partial_VS/\partial_VI$ obtained using $\partial_VS$ and
    $\partial_VI$ from the main plot. Panel ({\bf b}). $\partial_VI$ around
    $V=0$. As can be seen, at high $T$ the zero bias resonance in
    $\partial_VI$ is greatly suppressed below its universal maximum.}
\end{figure}

Notice, that when one is within the Majorana sweet spot region at low bias
voltages $|eV|\ll\Gamma$ at points with $|\eta_n|\neq|\eta_a|$ or at high bias
voltages $\sqrt{|\eta_n|^2+|\eta_a|^2}>|eV|\gg\Gamma$ at the point
$|\eta_n|=|\eta_a|$ and its vicinity, both the differential shot noise and
conductance may be rather small having a monotonous unremarkable
behavior. As a consequence, their separate analysis would not provide any
useful information about the physical nature of the states yielding the major
contribution to quantum transport behavior of the minimal Kitaev chain. For
example, Fig. \ref{figure_4}(a) demonstrates such a situation arising at high
bias voltages. Obviously, measuring only one physical quantity, $\partial_VS$
(black curve) or $\partial_VI$ (red curve), it is absolutely impossible to
conclude whether Majorana degrees of freedom are present or absent in the
system. In contrast, measuring both the differential shot noise and
differential conductance at high bias voltages, one obtains the ratio
$\partial_V S/\partial_V I$ which clearly demonstrates that exactly the poor
man's MBSs govern nonequilibrium response of the system fractionalizing the
differential effective charge to its universal value $q=3e/2$ as one can see
in the inset of Fig. \ref{figure_4}(a). The identification of the poor man's
MBSs via measurements of the fractional value $q=3e/2$ using the high voltage
tails of $\partial_VS$ and $\partial_VI$ has certain advantages over
measurements of the differential conductance around zero bias voltage. Here,
the differential conductance $\partial_VI$ is expected to have a resonance
when one tunes the system close to the center of the Majorana sweet spot
region, $|\eta_n|\approx|\eta_a|$. For poor man's MBSs the universal maximum
of this resonance is well known. Specifically, $\partial_VI=2e^2/h$ for the
configuration where one contact is coupled to QD1 and the other one to QD2 or,
as in this work, $\partial_VI=e^2/2h$ for the configuration where both
contacts are coupled to the same QD (for example, QD1 as in
Fig. \ref{figure_1}). However, since experiments are performed at finite
temperatures, maximal values of $\partial_VI$ at $V=0$ will be significantly
below the predicted universal maximum as demonstrated in
Fig. \ref{figure_4}(b). Thus at finite temperatures for a given configuration
one would be able to demonstrate only the existence of a resonance in
$\partial_VI$ at zero bias voltage but would discover that its maximal value
\begin{figure}
  \includegraphics[width=8.0 cm]{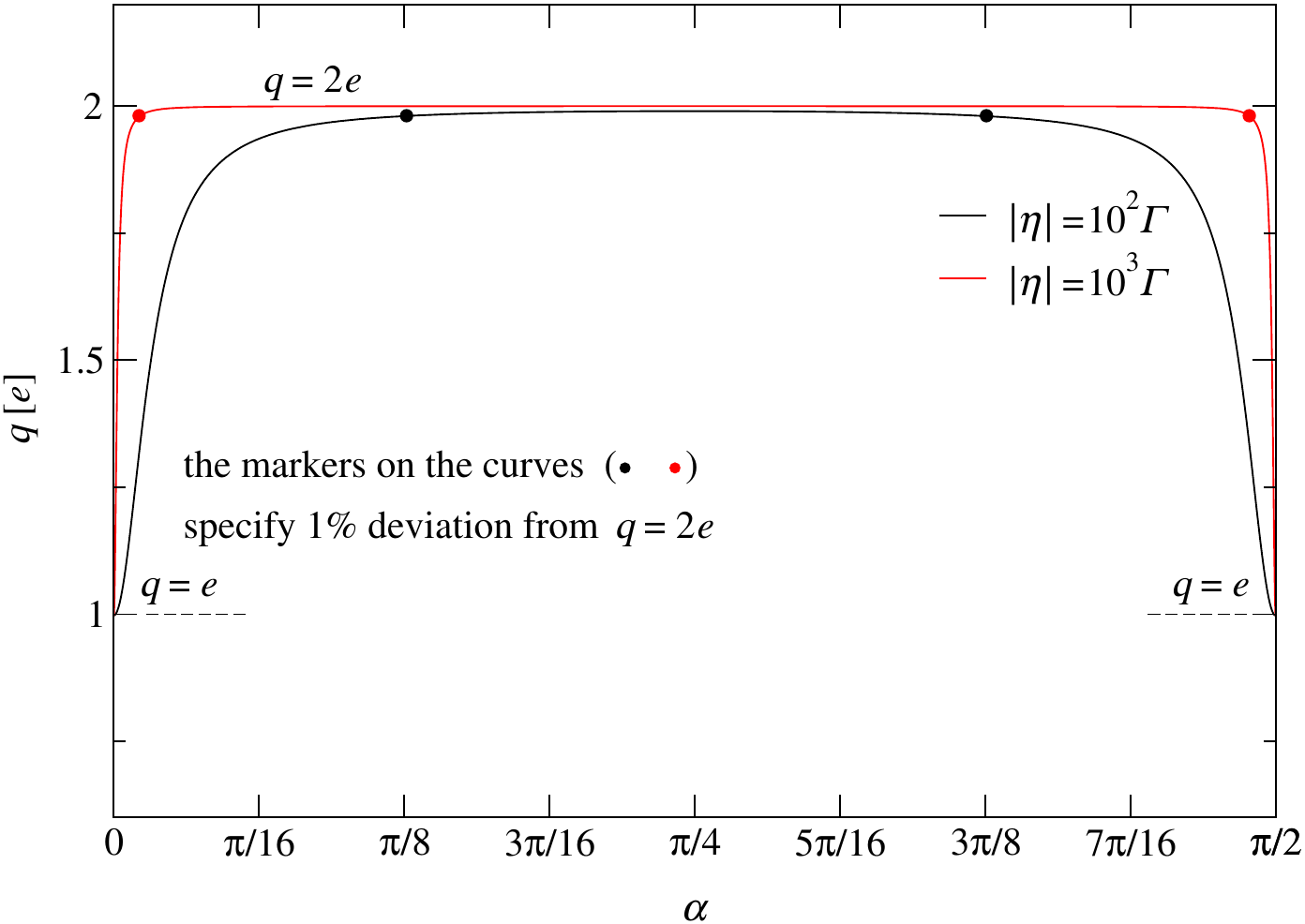}
  \caption{\label{figure_5} The differential effective charge,
    $q\equiv\partial_VS/\partial_VI$, is shown in universal units of the
    elementary charge $e$ as a function of the ratio between the amplitudes
    $|\eta_n|$ and $|\eta_a|$ for $|eV|=2\sqrt{|\eta_n|^2+|\eta_a|^2}$. Here
    $|\eta_n|=|\eta|\cos\alpha$, $|\eta_a|=|\eta|\sin\alpha$ with
    $|\eta|/\Gamma=10^2$ (black curve) and $|\eta|/\Gamma=10^3$ (red
    curve). With this parameterization the bias voltage does not vary along
    the curves, $|eV|=2|\eta|$. The values of the other parameters are the
    same as in Fig. \ref{figure_2}.}
\end{figure}
is below the exact universal maximum expected for poor man's MBSs. In fact,
this is observed in many experiments on minimal Kitaev chains. For example, in
Refs. \cite{Dvir_2023,Zatelli_2024,Haaf_2024,Bordin_2025,Haaf_2025} the
experiments are performed for the configurations where one expects the
differential conductance to reach the universal maximum $2e^2/h$ at $V=0$. All
these experiments reveal the existence of the resonance in $\partial_VI$ at
$V=0$ but the maximum of this resonance turns out to be much below the
universal value $2e^2/h$ which should have been detected for a clear
demonstration of poor man's MBSs. At the same time, as can be seen in
Fig. \ref{figure_4}(b), the high voltage tails of the differential
conductance, namely its parts for $|eV|\gg k_\text{B}T$, remain unchanged when
the temperature increases. When these high voltage tails are combined with the
ones of $\partial_VS$, one obtains the fractional value of the differential
effective charge, $q=3e/2$, which is expected for poor man's MBSs and may be
observed even at very high temperatures as demonstrated in
Fig. \ref{figure_4}(a).

When the bias voltage becomes very large,
$|eV|\gg\sqrt{|\eta_n|^2+|\eta_a|^2}$, the Majorana sweet spot region ruins
and the differential effective charge acquires its trivial integer value
$q=e$ for any value of the ratio $|\eta_n|/|\eta_a|$. However, before this
happens, the differential effective charge takes another integer
value. Specifically, we find numerically that at the bias voltage
$|eV|=2\sqrt{|\eta_n|^2+|\eta_a|^2}$ the differential effective charge takes
its maximal value which turns out to be integer, namely $q=2e$. Using the
parameterization in Eq. (\ref{Param_eta_n_eta_a}) we show in
Fig. \ref{figure_5} that the differential effective charge $q=2e$ may emerge
when the ratio $|\eta_n|/|\eta_a|$ varies in a wide range around the point
$\alpha=\pi/4$. The size of this range depends on $|\eta|$. We choose 1\%
deviation from $q=2e$ to specify the boundaries (the black and red dots in
Fig. \ref{figure_5}) of the range where $q\approx 2e$. As can be seen, the
size of this range quickly grows when $|\eta|$ increases. For example, for
$|\eta|/\Gamma=10^3$ the differential effective charge $q=2e$ is observed
almost in the whole range of $\alpha$ except for very narrow vicinities of the
edge points, $\alpha=0$ ($|\eta_n|=|\eta|$, $|\eta_a|=0$) and $\alpha=\pi/2$
($|\eta_n|=0$, $|\eta_a|=|\eta|$), where, as discussed above, the differential
effective charge takes its trivial integer value $q=e$ for any bias
voltage. We see that the Majorana sweet spot region has completely
disappeared. This is indicated by the differential effective charge whose
fractional value $q=3e/2$ has been fully washed out. However, before the
differential effective charge goes to its trivial integer value $q=e$ in the
whole range of $|\eta_n|/|\eta_a|$ for $|eV|\gg\sqrt{|\eta_n|^2+|\eta_a|^2}$,
there develops a non-Majorana region with $q=2e$ around the point
$|\eta_n|=|\eta_a|$ for the bias voltage
$|eV|=2\sqrt{|\eta_n|^2+|\eta_a|^2}$.

To qualitatively explain the fractional and integer values of the differential
effective charge $q$, one may try to associate fluctuations of the electric
current with fluctuations between the even and odd states of the minimal
Kitaev chain. As mentioned above (see Sec. \ref{intro}), poor man's MBSs
emerge when one tunes the system close to the even-odd degeneracy of its
ground state. In the neighborhood of this degeneracy point the electric
current fluctuates in response to various transitions between the even and odd
sectors. Such transitions are accurately captured by the shot noise and are
eventually quantified via a certain value of the differential effective
charge. When the minimal Kitaev chain is within its Majorana sweet spot region
and the bias voltage satisfies Eq. (\ref{Loc_of_e_half_high_V}), the
fluctuations between the even and odd sectors are realized via individual
transitions of two types: (1) between the states with the populations 0 and 1
(transitions of type $0\leftrightarrow 1$) and (2) between the states with the
populations 1 and 2 (transitions of type $1\leftrightarrow 2$). Each of these
Majorana transitions induces individual fluctuations which are mapped via the
shot noise to the fractional differential effective charge $q=e/2$. When the
system is within the Majorana sweet spot region but the difference
$|\eta_n|-|\eta_a|$ does not satisfy Eq. (\ref{Loc_of_e_half_high_V}), the
ordinary fluctuations induced by the coupling to the normal contacts become
comparable to the fluctuations between the even and odd sectors and provide
the well known contribution of one elementary charge $e$ which adds to $e/2$
coming from the Majorana even-odd fluctuations. As a result, away from the
points specified by Eq. (\ref{Loc_of_e_half_high_V}) there develops the
fractional value $q=3e/2$ which characterizes the main body of the Majorana
sweet spot region. Far away from the Majorana sweet spot region one naturally
expects that the Majorana fluctuations, that is the fluctuations between the
even and odd sectors, become less efficient and the electric current
fluctuates mainly due to the coupling to the normal contacts resulting in the
trivial integer value $q=e$ perfectly reached at the points $|\eta_n|\neq 0$,
$|\eta_a|=0$ and $|\eta_n|=0$, $|\eta_a|\neq 0$. The same happens when the
bias voltage applied to the normal contacts becomes very large,
$|eV|\gg\sqrt{|\eta_n|^2+|\eta_a|^2}$. In this situation fluctuations of the
electric current are also determined solely by the coupling to the normal
contacts and one observes the trivial integer value $q=e$ both within and
outside the Majorana sweet spot region, that is for any value of the ratio
$|\eta_n|/|\eta_a|$. However, before this trivial regime is reached for
$|eV|\gg\sqrt{|\eta_n|^2+|\eta_a|^2}$, the fluctuations between the even and
odd sectors of the minimal Kitaev chain may still play a role for bias
voltages of the order of the interdot coupling energy,
$|eV|\sim\sqrt{|\eta_n|^2+|\eta_a|^2}$. As we have seen above, the
differential effective charge in this regime reaches its maximal value $q=2e$
at $|eV|=2\sqrt{|\eta_n|^2+|\eta_a|^2}$. This result suggests that for
$|eV|\sim\sqrt{|\eta_n|^2+|\eta_a|^2}$ the fluctuations excited by the
transitions of type $0\leftrightarrow 1$ and of type $1\leftrightarrow 2$ are
not individually captured by the shot noise as it happens for
$|eV|<\sqrt{|\eta_n|^2+|\eta_a|^2}$. Instead, these two types of transitions
are counted as a whole in the shot noise measurements so that their merged
fluctuation response reaches its maximum at
$|eV|=2\sqrt{|\eta_n|^2+|\eta_a|^2}$ and contributes one elementary charge
$e$ (obtained as $e/2+e/2$) to the differential effective charge. This
contribution combines with the elementary charge $e$ coming from the
fluctuations induced by the normal contacts and in total one detects the
integer value of the differential effective charge, $q=2e$, which, obviously,
does not allow to reveal the Majorana fractionalization of the electronic
degrees of freedom.
\begin{figure}
  \includegraphics[width=8.0 cm]{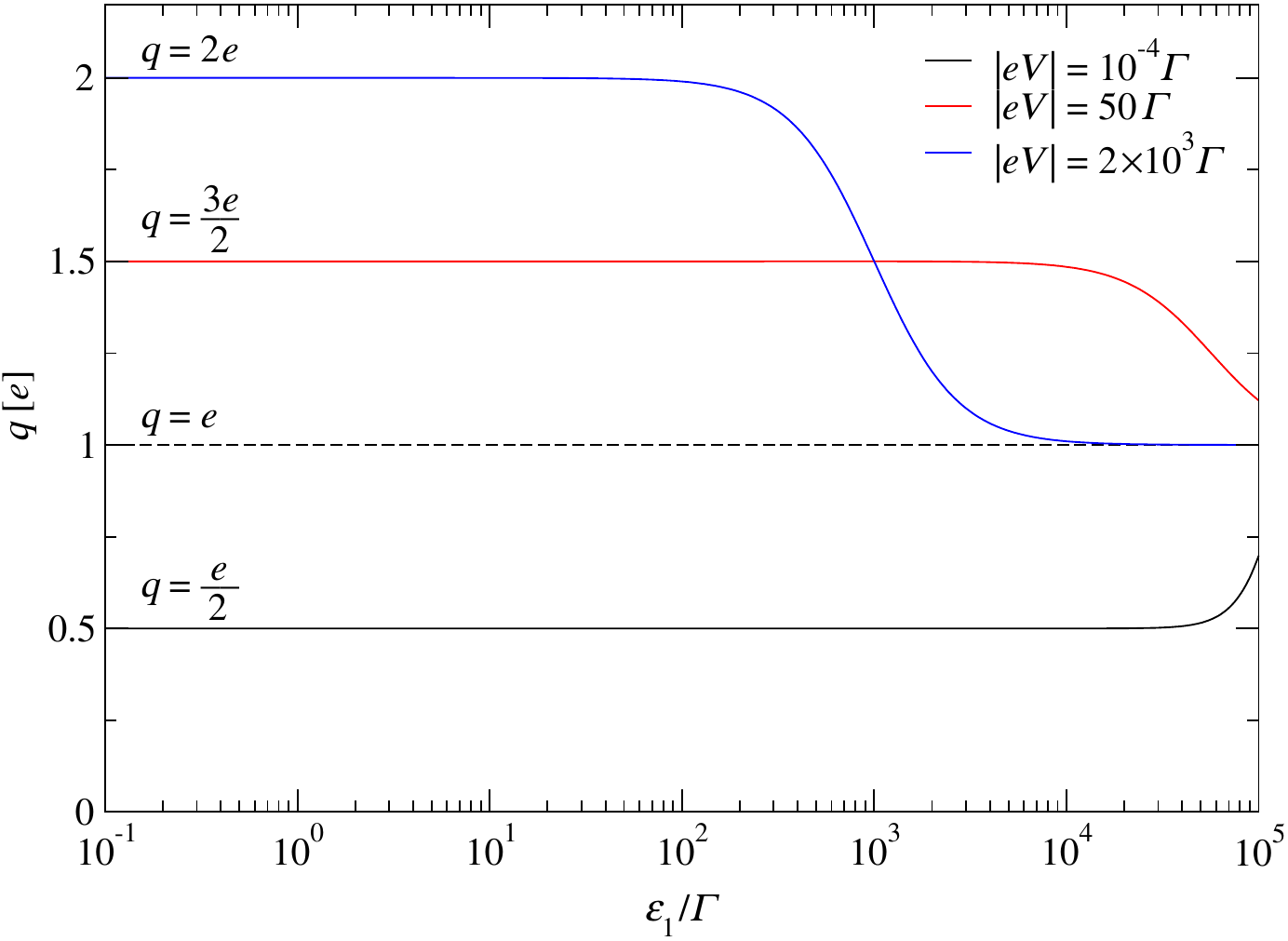}
  \caption{\label{figure_6} The differential effective charge,
    $q\equiv\partial_VS/\partial_VI$, is shown in universal units of the
    elementary charge $e$ as a function of the gate voltage on QD1, specified
    by $\varepsilon_1$. Here for all the curves $|\eta_n|=|\eta|\cos\alpha$,
    $|\eta_a|=|\eta|\sin\alpha$ with $|\eta|/\Gamma=10^3$ and $\alpha=\pi/4$
    (or $|\eta_n|=|\eta_a|$). The black curve is obtained in the regime
    $|eV|\ll\Gamma$, specifically for $|eV|/\Gamma=10^{-4}$. The red curve is
    obtained in the regime $\Gamma\ll|eV|<\sqrt{|\eta_n|^2+|\eta_a|^2}$,
    specifically for $|eV|/\Gamma=50$. The blue curve is obtained for
    $|eV|=2\sqrt{|\eta_n|^2+|\eta_a|^2}$, that is
    $|eV|/\Gamma=2|\eta|/\Gamma=2\times 10^3$. The other parameters have the
    same values as in Fig. \ref{figure_2}.}
\end{figure}

It is also important to consider how the differential effective charge depends
on the gate voltage applied to QD1 to tune $\varepsilon_1$. To obtain
essential characteristics of this dependence it is enough to use the
parameterization in Eq. (\ref{Param_eta_n_eta_a}). We find numerically that
within the universal Majorana regime specified in
Eq. (\ref{Univ_Majorana_reg}) the differential effective charge does not
depend on $\varepsilon_1$. In particular, as Fig. \ref{figure_6} demonstrates,
both Majorana fractional values, $q=e/2$ and $q=3e/2$, appearing at the point
$|\eta_n|=|\eta_a|$ for low (black curve) and high (red curve) bias voltages,
are quite stable when $\varepsilon_1$ is varied within the range
$|\varepsilon_1|<\sqrt{|\eta_n|^2+|\eta_a|^2}$. Moreover, the black and red
curves demonstrate that the Majorana sweet spot region is very robust against
large values of $\varepsilon_1$ and the poor man's MBSs continue to drive the
fluctuation behavior of the minimal Kitaev chain even when the gate voltage
applied to QD1 is outside the universal Majorana regime defined in
Eq. (\ref{Univ_Majorana_reg}). Indeed, the black and red curves start to
slightly deviate from the Majorana fractional values $q=e/2$ and $q=3e/2$
towards the trivial integer value $q=e$ when $\varepsilon_1$ is already well
above the upper bound of the universal Majorana regime,
$\varepsilon_1\gg\sqrt{|\eta_n|^2+|\eta_a|^2}$. In contrast, the non-Majorana
integer value $q=2e$ is less robust against large values of
$\varepsilon_1$. This is demonstrated in Fig. \ref{figure_6} by the blue curve
which exhibits a strong dependence on $\varepsilon_1$ already within the
universal Majorana regime, $|\varepsilon_1|<\sqrt{|\eta_n|^2+|\eta_a|^2}$, and
reaches the value $q=e$ long before the black and red curves converge to this
trivial integer limit.
\begin{figure}
  \includegraphics[width=8.0 cm]{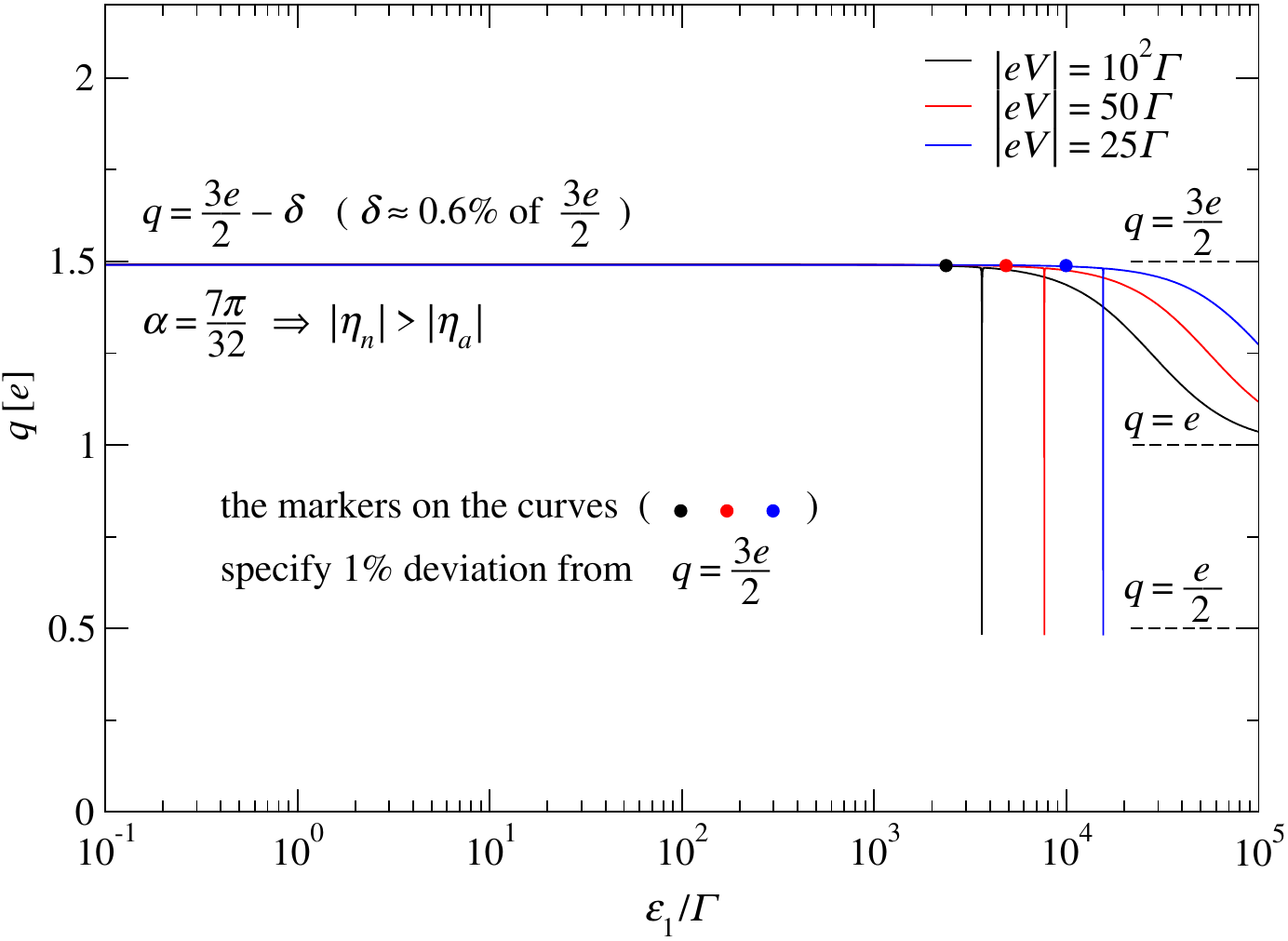}
  \caption{\label{figure_7} The differential effective charge,
    $q\equiv\partial_VS/\partial_VI$, is shown in universal units of the
    elementary charge $e$ as a function of the gate voltage on QD1, specified
    by $\varepsilon_1$, for different values of the bias voltage in the
    strongly nonequilibrium regime,
    $\Gamma\ll|eV|<\sqrt{|\eta_n|^2+|\eta_a|^2}$. Here for all the curves
    $|\eta_n|=|\eta|\cos\alpha$, $|\eta_a|=|\eta|\sin\alpha$ with
    $|\eta|/\Gamma=10^3$ and $\alpha=7\pi/32$ (or $|\eta_n|>|\eta_a|$). The
    bias voltage is $|eV|/\Gamma=10^2$ (black curve), $|eV|/\Gamma=50$ (red
    curve), $|eV|/\Gamma=25$ (blue curve). The values of the other parameters
    are the same as in Fig. \ref{figure_2}.}
\end{figure}

Finally, we analyze the dependence of the differential effective charge on the
gate voltage applied to QD1 for the case $|\eta_n|\neq|\eta_a|$ when the
minimal Kitaev chain is within the Majorana sweet spot region for small gate
voltages. In Fig. \ref{figure_7} we present the results obtained for
$\alpha=7\pi/32$ which, according to the parameterization in
Eq. (\ref{Param_eta_n_eta_a}), means that $|\eta_n|>|\eta_a|$. In this case
the differential effective charge is not exactly equal to the Majorana
fractional value $q=3e/2$ but it is a bit smaller. Specifically, for the
chosen value of $\alpha$ and within the universal Majorana regime,
$|\varepsilon_1|<\sqrt{|\eta_n|^2+|\eta_a|^2}$, the differential effective
charge is 0.6\% smaller than the Majorana fractional value $q=3e/2$. Thus the
chosen value of $\alpha$ is within the Majorana sweet spot region whose
boundaries have been defined above as the points with 1\% deviation from the
fractional value $q=3e/2$. The curves in Fig. \ref{figure_7} obtained for
different bias voltages demonstrate that the differential effective charge
does not depend on the bias and gate voltages within the universal Majorana
regime. This illustrates stability of the Majorana sweet spot region when both
the bias and gate voltages are varied within the range
$|eV|,|\varepsilon_1|<\sqrt{|\eta_n|^2+|\eta_a|^2}$. For larger gate voltages,
$\varepsilon_1>\sqrt{|\eta_n|^2+|\eta_a|^2}$, the differential effective
charge acquires dependence on $\varepsilon_1$ and starts to decrease to its
trivial integer value $q=e$. The black, red and blue dots on the corresponding
curves label where the system exits the Majorana sweet spot region, that is
where the differential effective charge becomes 1\% smaller than the
fractional value $q=3e/2$. We see that all the dots are located outside the
universal Majorana regime, $\varepsilon_1>\sqrt{|\eta_n|^2+|\eta_a|^2}$, and
that the gate voltage, at which the system exits its Majorana sweet spot
region, gets lower and lower when the bias voltage gradually
increases. Further, as can be seen in Fig. \ref{figure_7}, soon after the
minimal Kitaev chain quits its Majorana sweet spot region, it passes through a
very narrow region where the differential effective charge takes its minimal
value. This very narrow region is similar to those observed in
Figs. \ref{figure_2} and \ref{figure_3}, where these narrow minima indicate
locations of the resonances in the differential shot noise and conductance
with the Majorana fractional ratio $q=e/2$ at the central points of the
resonances. Here, however, these resonances occur a bit outside the Majorana
sweet spot region and thus their ratio turns out to be approximately 3.3\%
smaller than the Majorana fractional value $q=e/2$. Within this work we do not
derive the exact expression for the gate voltage $\varepsilon_1^*$ around
which the narrow minima, observed in Fig. \ref{figure_7}, are
located. However, numerically we identify the main contribution to
$\varepsilon_1^*$ as
\begin{equation}
  \varepsilon_1^*\approx 2\frac{|\eta_n|^2-|\eta_a|^2}{|eV|}.
  \label{eps1_star}
\end{equation}
The precision of this approximation gradually increases when the value of its
right hand side gets larger and larger, for example, when the bias voltage
decreases, as demonstrated in Fig. \ref{figure_7}.
\section{Conclusion}\label{Conclusion}
In this work we have explored nonequilibrium fluctuation behavior of a minimal
Kitaev chain composed of two QDs coupled via normal tunneling and crossed
Andreev reflection characterized by the amplitudes $|\eta_n|$ and
$|\eta_a|$. To quantitatively describe the fluctuation response of the
electric currents in this system we have introduced the differential effective
charge $q$ defined as the ratio, $q\equiv\partial_V S/\partial_V I$, of the
differential shot noise and conductance. We have numerically calculated the
differential effective charge as a function of the ratio $|\eta_n|/|\eta_a|$
and gate voltage, applied to one of the QDs, for both low and high bias
voltages $|eV|$. In the regime of low bias voltages it has been found that the
Majorana sweet spot region of the weakly nonequilibrium minimal Kitaev chain
is characterized by two fractional values of the differential effective
charge, specifically, $q=e/2$ and $q=3e/2$. The fractional value $q=e/2$
arises in a very narrow vicinity of the point $|\eta_n|=|\eta_a|$ whereas the
fractional value $q=3e/2$ emerges in the main body of the Majorana sweet spot
region. Thus in a weakly nonequilibrium minimal Kitaev chain the fractional
value $q=3e/2$ may be chosen as an indicator of the Majorana sweet spot
region. Outside the Majorana sweet spot region the differential effective
charge monotonously decreases and reaches its trivial integer value $q=e$ at
the points ($|\eta_n|\neq 0$, $|\eta_a|=0$) and ($|\eta_n|=0$,
$|\eta_a|\neq 0$). In fact, at these points the differential effective charge 
is independent of the bias voltage and retains its trivial integer value
$q=e$. In the regime of high bias voltages we have found that the main body of
the Majorana sweet spot region does not change and is still characterized by
the fractional value $q=3e/2$. However, the very narrow region with $q=e/2$
disappears from the point $|\eta_n|=|\eta_a|$ where the differential effective
charge now fractionalizes to $q=3e/2$. In contrast to the weakly
nonequilibrium situation, now there appear two narrow regions with $q=e/2$
around the points $|\eta_n|-|\eta_a|=\pm|eV|/2$. Except for the very narrow
domains with $q=e/2$ the Majorana sweet spot regions for low and high bias
voltages coincide and are characterized by the fractional value $q=3e/2$. For
very large bias voltages, $|eV|\gg\sqrt{|\eta_n|^2+|\eta_a|^2}$, we have found
that the Majorana sweet spot region is completely destroyed as can be inferred
from the differential effective charge which takes its trivial integer value
$q=e$ for any value of the ratio $|\eta_n|/|\eta_a|$. However, before this
trivial state is reached, the differential effective charge increases up to
its maximal value which turns out to be integer. More precisely, our numerical
calculations show that for the bias voltage
$|eV|=2\sqrt{|\eta_n|^2+|\eta_a|^2}$ the differential effective charge takes
the integer value $q=2e$ in a wide range of the ratio $|\eta_n|/|\eta_a|$ and
that this range quickly widens when the interdot coupling energy
$\sqrt{|\eta_n|^2+|\eta_a|^2}$ increases. As a consequence, at large values of
the interdot coupling energy the non-Majorana integer value $q=2e$ is observed
almost in the whole range of the ratio $|\eta_n|/|\eta_a|$ except for very
narrow vicinities of the points ($|\eta_n|\neq 0$, $|\eta_a|=0$) and
($|\eta_n|=0$, $|\eta_a|\neq 0$) where, as mentioned above, $q=e$. Finally,
analyzing how the differential effective charge depends on the gate voltage,
it has been demonstrated that the Majorana sweet spot region is quite robust
against large gate voltages. In particular, the fractional value $q=3e/2$,
characterizing the main body of the Majorana sweet spot region, persists
within and even to some extent outside the universal Majorana regime as has
been verified at various points of the Majorana sweet spot region: at the
central point, where $|\eta_n|=|\eta_a|$, and at a general point, where
$|\eta_n|\neq|\eta_a|$. In contrast, the non-Majorana integer value $q=2e$ is
not stable and ruins for much smaller gate voltages already within the
universal Majorana regime.

The results presented here for a double QD Kitaev chain demonstrate its
remarkable fluctuation behavior and that fluctuations of electric currents in
various short Kitaev chains may in general exhibit potentially rich
behavior. This behavior is not only of fundamental interest but also of
practical importance. In particular, our results suggest that the Majorana
fractional value of the differential effective charge, $q=3e/2$, represents a
reliable indicator of the Majorana sweet spot region both for weakly and
strongly nonequilibrium minimal Kitaev chains. It may be used to estimate the
quality of poor man's MBSs as an experimentally appealing alternative to the
Majorana polarization. This is relevant in situations where the Majorana
polarization is close to unity but the even-odd degeneracy of the system's
ground state is absent \cite{Tsintzis_2022}. Indeed, in these situations the
differential effective charge would obviously take an integer value indicating
the absence of Majorana degrees of freedom. Further, this indicator might also
be useful when separate measurements of the shot noise and conductance are
performed within the Majorana sweet spot region but yield small values or
non-resonant behavior having no special features and providing no useful
information about whether the minimal Kitaev chain is within its Majorana
sweet spot region or not. This might happen for strongly nonequilibrium
minimal Kitaev chains and also in the weakly nonequilibrium regime when one is
within the Majorana sweet spot region but $|\eta_n|\neq|\eta_a|$. In all such
cases, the ratio of the differential shot noise and conductance, that is the
differential effective charge, fractionalizes to $q=3e/2$ and, in contrast to
separate measurements, clearly reveals that the poor man's MBSs drive the
fluctuation behavior of the minimal Kitaev chain. Moreover, it has been
demonstrated that measurements of the fractional value $q=3e/2$ are feasible
even at high temperatures. This provides a great advantage over experiments
measuring only differential conductances. Indeed, such experiments have been
able to detect the zero bias resonance in the differential
conductance. However, a clear demonstration that the maximum of this resonance
reaches exactly the universal value predicted for poor man's MBSs is difficult
because experiments are performed at finite temperatures which strongly
suppress the maximum of the zero bias resonance in the differential
conductance. Thus in many experiments
\cite{Dvir_2023,Zatelli_2024,Haaf_2024,Bordin_2025,Haaf_2025} one observes the
zero bias resonance in the differential conductance but the maximum of this
resonance does not reach the universal Majorana value which should have been
detected for a clear demonstration of poor man's MBSs. In contrast, the
differential effective charge may be obtained from the high voltage tails of
the differential shot noise and conductance. These tails are robust against
high temperatures and may be used to measure the fractional value $q=3e/2$
predicted for the differential effective charge of poor man's MBSs even at
high temperatures. Since successful conductance measurements have already been
performed, we hope that both the fundamental appeal and practical relevance of
fluctuation phenomena governed by poor man's MBSs will stimulate the next
generation of experimental research on nonequilibrium shot noise in minimal
Kitaev chains.

\section*{Acknowledgments}
The author thanks Reinhold Egger for an important discussion and comments.

\end{document}